\documentclass[fleqn,twoside]{article}
\usepackage{espcrc2}

\usepackage{graphicx}
\usepackage{amsmath}
\usepackage{amssymb}

\usepackage[figuresright]{rotating}

\newcommand{\eg}{{\it e.g.}}

\newcommand{\eq}{Eq.}

\newcommand{\fig}{Fig.}

\newcommand{\Ref}{Ref.}
\newcommand{\Refs}{Refs.}

\newcommand{\stheta}{\sin^22\theta_{13}}
\newcommand{\deltacp}{\delta_\mathrm{CP}}

\newcommand{\equ}[1]{\eq~(\ref{equ:#1})}
\newcommand{\figu}[1]{\fig~\ref{fig:#1}}

\newcommand{\bi}{\begin{itemize}}
\newcommand{\ei}{\end{itemize}}

\title{Neutrino oscillation physics with a FNAL proton driver}

\author{Walter Winter\address{School of Natural Sciences, Institute for Advanced Study,
        Princeton, NJ 08540}%
        \thanks{Work supported by the W.~M.~Keck Foundation and NSF grant PHY-0070928.}}

\begin{document}

\begin{abstract}
We discuss the need of a proton driver for the Fermilab neutrino
oscillation program, as well as its role in the global context.
%\vspace{-1pc}
\end{abstract}

\maketitle

%\section{Introduction}

The Fermilab proton driver (FPD) is proposed to consist of a linear accelerator and
several main injector modifications. Its purpose is to provide higher fluxes to potential experiments, such as neutrino oscillation experiments~\cite{PDUD,PDNOD}, or
muon, kaon, pion, neutron, or antiproton experiments~\cite{PDUD}. It is aiming for $8 \, \mathrm{GeV}$ and (up to) $120 \, \mathrm{GeV}$ protons at about $2 \, \mathrm{MW}$, which is a factor of five to ten past the current FNAL proton source at $120 \, \mathrm{GeV}$.
A proton driver is one of the most important prerequisites for a long-term neutrino oscillation program at Fermilab, because any currently planned long-baseline experiment is limited by the luminosity $\mathcal{L}$:
\begin{equation}
\mathcal{L} = \mathrm{Flux} \times \mathrm{Detector \, mass} \times \mathrm{Running \, time}
\label{equ:l}
\end{equation}
For example, the $\stheta$ sensitivity at the NO$\nu$A superbeam~\cite{Ayres:2004js}
is, to a first order, proportional to $1/\sqrt{\mathcal{L}}$. Thus, a factor of two better
$\stheta$ sensitivity requires about a factor of four in the luminosity, and one is quickly running
into limitations if one wants to obtain significantly better sensitivities. From \equ{l},
there are three possibilities for such a luminosity upgrade of about a factor of five:
\begin{itemize}
\item
A flux upgrade, where the factor of five corresponds to the FPD
\item
Longer running times, where a factor of five corresponds to about $25$ years
\item
Larger or more efficient detectors, where a factor of five in size corresponds to a $150 \, \mathrm{kt}$ liquid scintillator for NO$\nu$A
\end{itemize}
Since such long running times are unrealistic, and larger detectors might not only be more
expensive\footnote{There could be, however, alternatives, such as a considerably smaller liquid argon detector~\cite{Bartoszek:2004si}. In this case, even much higher luminosities could be reached together with the FPD (see, \eg, \Ref~\cite{Mena:2005sa}).}, but also in most cases not be re-usable for new experiments, the FPD
option is certainly a promising investment. Especially, the potential of the FPD to be used for successive generations of experiments makes it a key
component in a long-term neutrino oscillation program. Potential users
include MiniBOONE, NO$\nu$A, Super-NO$\nu$A (NO$\nu$A with a 2nd detector), a broad band superbeam with a new beamline, a $\beta$-Beam, a neutrino factory, and many other options (see \Ref~\cite{PDNOD} and references therein).
Thus, there is a number of different experiments with a beam based on the FPD.
Which of these experiments should be built, however, depends on the physics case.
Therefore, the key questions for the FPD in the context of the neutrino oscillation program are:
\begin{enumerate}
\item
 What are the relevant physics scenarios?
\item
 Does one need the FPD in {\em any} physics case?
\item
 What does one do with the protons then?
\item
 Why build the proton driver at Fermilab?
\end{enumerate}
These questions should be answered by the neutrino oscillation part of the
FPD study~\cite{PDNOD}, which therefore has a quite strategical character.

%\section{Performance indicators}
% MAYBE LATER OR OMIT

%\section{Physics scenarios}

In order to lay out the physics scenarios, we need to understand the pre-FPD
neutrino oscillation program. For the FPD, operation is supposed to start around 2014.
Thus, there are about ten more years to go until then.  The potential of future
reactor (Double Chooz and larger ones) and accelerator (MINOS, CNGS, T2K, and NO$\nu$A) neutrino oscillation experiments for the coming ten years has, for example, been studied in \Ref~\cite{Huber:2004ug}. For the FPD case, we could identify the following three main physics scenarios with respect to the actual value of $\stheta$~\cite{PDNOD}:
\begin{description}
\item[Scenario 1] $\stheta \gtrsim 0.04$. In this case, $\stheta$ will have been certainly
discovered before FPD startup. As a consequence, one could start studying the neutrino mass hierarchy and CP violation already with existing beamline and detector(s) using the
FPD.
\item[Scenario 2] $0.01 \lesssim \stheta \lesssim 0.04$. In this scenario, $\stheta$ is likely
(depending on $\deltacp$ and mass hierarchy) to be discovered before FPD startup.
For substantial mass hierarchy and CP violation sensitivities, upgrades beyond the existing beamline and detector are necessary. However, superbeams based upon the FPD
are the appropriate technology choice.
\item[Scenario 3] $\stheta \lesssim 0.01$. A $\stheta$ discovery is unlikely before proton
driver startup, and a neutrino factory (or $\beta$-Beam) program might be required for
the sensitivities to $\stheta$, the mass hierarchy, and CP violation. The FPD can be used to obtain better $\stheta$ limits, and as a component of such a program.
\end{description}
Though the numbers chosen to separate these three cases somewhat depend on assumptions,
definitions of the performance indicator, and confidence level, it important to keep in mind that
these three physics cases represent three very conceptually different Fermilab neutrino oscillation program alternatives. Below, we will discuss these scenarios in greater detail.

Except from these three main scenarios, we have identified three special cases which could be especially interesting for the FPD (for details, see \Ref~\cite{PDNOD}):
\begin{description}
\item[Special case 1] $\theta_{23}$ still consistent with maximal mixing. Because maximal atmospheric mixing is an important indicator for neutrino mass models (such as it could be a hint for a flavor symmetry), small deviations from maximal mixing should be tested. One possibility is
NO$\nu$A together with the FPD.
\item[Special case 2] LSND confirmed. Among other possibilities, new short baseline experiments could test the contribution of sterile neutrinos or $\nu_\mu$ to $\nu_\tau$ transitions
using the FPD.
\item[Special case 3] Something else unexpected happens. In this special case, higher
luminosities, such as from the FPD, are almost certainly required to study the nature of the unexpected effect.
\end{description}
In all of the discussed main scenarios and special cases, the FPD is the important key component to obtain higher luminosities. Thus, the FPD
is very useful in any physics case.

%\section{What to do with the protons}

Let us now discuss what to do with the protons in the three main $\stheta$ scenarios. In particular, we show several possibilities for FPD-based experiments to address the relevant
remaining questions of mass hierarchy and CP violation.

For Scenario~1 ($\stheta \gtrsim 0.04$), a mass hierarchy determination would be possible for a substantial fraction of the parameter space by the combination of the FPD with possibly existing equipment, such as NO$\nu$A. In addition, for both measurements, a real synergy with the Japanese T2K-program can be found -- provided that T2K gets an upgraded proton driver, too.
If T2K did not get any upgrades (proton driver or detector), the measurements could be done at Fermilab, too, such as in Scenario~2 (below).

In Scenario~2 ($0.01 \lesssim \stheta \lesssim 0.04$), a second detector in the NuMI beamline
together with the FPD would cover most of the parameter space for both the mass hierarchy and CP violation measurements (see, \eg, \Ref~\cite{MenaRequejo:2005hn}). In addition, a new beamline broad band beam using the FPD, such as with a long baseline targeted towards a deep underground laboratory, would have an excellent potential for both measurements even compared to the proposed T2HK program. Several options are currently studied and discussed (see \Ref~\cite{PDNOD}).

Eventually, for Scenario~3 ($\stheta \lesssim 0.01$), a neutrino factory program could
find $\stheta$ and determine the neutrino mass hierarchy and CP violation at least down to $\stheta \sim 10^{-4}$~\cite{Huber:2003ak}. In addition, a higher gamma $\beta$-Beam might be a possible alternative~\cite{Burguet-Castell:2003vv,Huber:2005jk}. For both programs,
a FPD would be a useful component. Note, however, that
one part of this parameter space ($\stheta \gtrsim 0.005$) can also be probed by the new
beamline experiment from Scenario~2. In fact, one can show that especially in this range no proton driver at Fermilab might lead to a hold of the global superbeam programs because $\stheta$
might be believed to be too small from T2K and NO$\nu$A alone. However, in this case, all of the interesting measurements could be probably done with superbeams, which means that a great
opportunity would be missed.

%\section{The FPD in the global context}

\begin{figure}
\begin{center}
\includegraphics[width=\columnwidth]{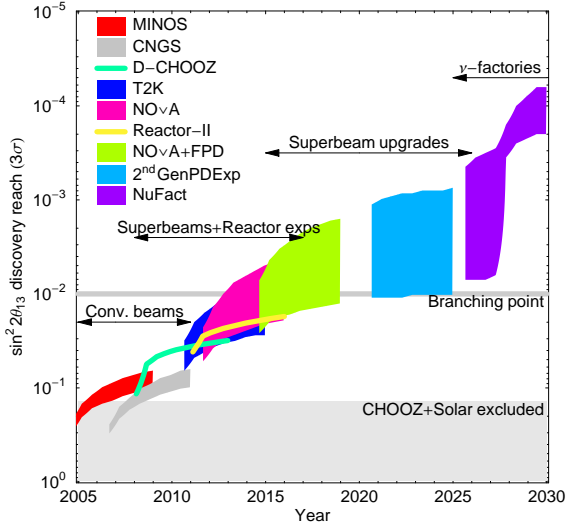}
\vspace*{-1.5cm}
\end{center}
\caption{\label{fig:bands} Possible global evolution of the $\stheta$
discovery reach. Figure taken from \Refs~\cite{PDUD,PDNOD}.}
\end{figure}

We show in \figu{bands} one possible global evolution of the $\stheta$ discovery potential
for the normal mass hierarchy, where the dependence on the actual value of the unknown CP
phase is illustrated as bands (for details, see \Ref~\cite{PDNOD}). In this figure,
(potential) Fermilab-based experiments are MINOS, NO$\nu$A, NO$\nu$A+FPD,
a 2nd generation proton driver experiment (2$^{\mathrm{nd}}$GenPDExp), and the neutrino factory (NuFact). Obviously, the FPD is needed for the ``natural'' extension of the current Fermilab
neutrino oscillation program. In addition, the FPD-based $\stheta$ discovery potential would likely exceed any of the currently existing or planned experiments including a large reactor experiment. Note that the fact that the reactor experiments are not affected by $\deltacp$ also implies the complementarity to the beam experiments: They do not provide any information on $\deltacp$. An important argument for the proton driver at Fermilab is, of course, the existing NuMI beamline, which could lead to a continuous program as function of time. This beamline is complementary to the Japanese T2K program, too, because a mass hierarchy determination
requires a substantially longer baseline than currently planned for T2K. All these facts together with the present expertise from the running MiniBOONE and MINOS experiments make Fermilab a unique potential neutrino oscillation laboratory. Because we need the FPD in all relevant physics cases,
we conclude that the FPD is the next logical step in the evolution of the neutrino oscillation program at Fermilab beyond NO$\nu$A.

%\bibliographystyle{h-elsevier}
%\bibliography{references}

\end{document}